\documentclass[a4paper,11pt]{article}
\pdfoutput=1

\usepackage{jinstpub}

\usepackage{graphicx}
\usepackage{hyperref}
\usepackage{siunitx}
\usepackage[frozencache,cachedir=pygments]{minted}

\def\fwn{\texttt{toise}} %fwn = framework name

\def\figureautorefname{Fig.}

\def\sectionautorefname{Sec.}

\begin{document}

\title{\fwn: a framework to describe the performance of high-energy neutrino detectors}

\author[a]{J.~van Santen}
\author[b]{B.~A.~Clark}
\author[b]{R.~Halliday}
\author[a]{S.~Hallmann}
\author[a,c]{A.~Nelles}

\affiliation[a]{Deutsches Elektronen-Synchrotron DESY, Platanenallee 6, 15738 Zeuthen, Germany}
\affiliation[b]{Dept. of Physics and Astronomy, Michigan State University, East Lansing, MI 48824, USA}
\affiliation[c]{ECAP, Friedrich-Alexander-Universit{\"a}t Erlangen-N{\"u}rnberg, 91058 Erlangen, Germany}

\emailAdd{jakob.van.santen@desy.de}

\abstract{
Neutrinos offer a unique window to the distant, high-energy universe. Several next-generation instruments are being designed and proposed to characterize the flux of TeV--EeV neutrinos. 
The projected physics reach of the detectors is often quantified with simulation studies.
However, a complete Monte Carlo estimate of detector performance is costly from a computational perspective, restricting the number of detector configurations considered when designing the instruments. 
In this paper, we present a new Python-based software framework, \fwn{}, which forecasts the performance of a high-energy neutrino detector using parameterizations of the detector performance, such as the effective areas, angular and energy resolutions, etc. 
The framework can be used to forecast performance of a variety of physics analyses, including sensitivities to diffuse fluxes of neutrinos and sensitivity to both transient and steady state point sources.
This parameterized approach reduces the need for extensive simulation studies in order to estimate detector performance, and allows the user to study the influence of single performance metrics, like the angular resolution, in isolation.  
The framework is designed to allow for multiple detector components, each with different responses and exposure times, and supports paramterization of both optical- and radio-Cherenkov (Askaryan) neutrino telescopes.
In the paper, we describe the mathematical concepts behind \fwn{} and introduce the reader to the use of the framework.

%A complete Monte Carlo simulation of the performance of high-energy neutrino detectors is time-consuming both in set-up time of a certain detector configuration and the computing time itself. This then often restricts the number of configurations considered when designing detectors. The \fwn{} framework addresses these shortcomings by providing a parameterized approach to estimating typical aspects of the performance of neutrino detectors. This simplifies the design process and the study of the influence of single performance parameters without the overhead of a full Monte Carlo simulation. We introduce the mathematical concepts behind \fwn{} and provide detailed descriptions of instructive example applications.
}

\maketitle

\section{Introduction}

High-energy neutrinos are key messengers to the distant, high energy universe. To this end, several high energy neutrino experiments have been built in the last decade with the goal of discovering and characterizing the flux of TeV--EeV neutrinos. This includes experiments relying on the optical-Cherenkov light produced in neutrino-nucleon interactions, such as IceCube~\cite{IceCube:2016zyt}, ANTARES~\cite{ANTARES:2011hfw}, KM3NeT~\cite{KM3Net:2016zxf}, Baikal-GVD~\cite{Avrorin:2011zzc}, as well as experiments that rely on the radio-Cherenkov light produced by super-PeV neutrinos, such as RICE~\cite{Kravchenko:2011im}, ARIANNA~\cite{Anker:2019rzo}, and ARA~\cite{ARA:2019wcf}.
Analysis of data from these high-energy neutrino detectors requires a detailed model of the detector response that can be used to predict observations
under some physics scenario. This model typically has such a large number of dimensions that it is infeasible to evaluate directly, and instead must be estimated via Monte Carlo (MC) simulation. This can be quite computationally expensive, requiring hundreds to thousands of CPU- or GPU-hours per hour of detector livetime. For example, the IceCube collaboration reports that to achieve 10 years of simulated detector livetime with adequate statistics, they require 6M CPU hours for signal simulation and over $\sim30$k CPU \textit{years} for background simulation~\cite{IceCubeSnowmassLOI}.

When designing a future high-energy neutrino detector, like IceCube-Gen2 \cite{IceCube-Gen2:2020qha} or P-ONE \cite{Resconi:2021ezb}, it is infeasible to undertake these enormous simulation efforts while the designs are still rapidly evolving. Further, it can be helpful to estimate how a detector's projected sensitivity would compare to that of existing, well-understood analyses.  In this paper, we present a framework for producing such sensitivity estimates, using a simplified and factorized model of the detector response based on targeted MC simulations. This makes it possible to efficiently compare different detector designs without repeating the entire simulation chain, as well as study the influence of various factors on the detector's sensitivity in isolation.

The \fwn{} framework is written in Python \cite{Python3}, installable with conda \cite{install,anaconda,conda_forge_community_2015_4774217} and pip \cite{pip}, and utilizes Jupyter notebooks to illustrate its features and capabilities. We provide comprehensive tutorials that also form the basis of this article. 

\section{Describing a neutrino detector}
\label{sec:description_intro}

A neutrino detector can be fully described by a few functions. While accurately obtaining each of these may be difficult, they are straightforward from a didactic point of view. The first is a model for the neutrino flux and its relevant interactions in the sensitive volume. 
In \fwn, this is parameterized as a transfer tensor between initial neutrino flavor states and observable final states like muons, and is described in \autoref{sec:neutrino_physics}. 
The second is the acceptance of the detector to signals and backgrounds, which are the analysis/selection efficiencies and the effective areas; this is described in \autoref{sec:a_eff}. The third and fourth components are the detector's angular and energy resolution with respect to the observable final states, and are discussed in \autoref{sec:detector_resolution}.

In the \fwn{} framework, the full information on an instrument's response is encoded in a five-dimension response tensor $D_{f,E_\mathrm{\nu},\cos(\theta_\nu),E_\mathrm{rec},\Delta\Psi}$. The dimensions of this response tensor are the six neutrino types $f$ (3 (anti-)neutrino flavors), the true neutrino energy $E_\nu$ and direction $\cos(\theta_\nu)$, the reconstructed energy $E_\mathrm{rec}$ and the angular reconstruction uncertainty $\Delta\Psi$. This section describes both the general approach, as well as how the parameterizations are treated within the framework.

\subsection{Neutrino Physics }
\label{sec:neutrino_physics}

Fluxes of atmospheric and astrophysical neutrinos are typically parameterized at the surface of the Earth, but a neutrino detector below the surface is sensitive only to the number of neutrino-induced leptons that reach the instrumented volume and the number of neutrino-induced particle showers (also referred to as \emph{cascades}) that occur inside the sensitive volume. In order to separate the part of the detector performance that can be influenced by design choices from the limitations imposed by the physics of neutrino interactions, we split the event rate calculation into two stages: neutrino physics and detection. In the neutrino physics stage described in this section, we convert neutrino fluxes at the surface of the Earth to area or volume densities of final states at the detector. These densities serve as input to the detection stage described in \sectionautorefname~\ref{sec:a_eff} and do not depend on the detector configuration.

In the neutrino energy range relevant for optical neutrino detection, two general classes of neutrino-induced events are relevant. The first are incoming muon events, where a neutrino interaction far from the detector produces a high-energy muon that reaches the instrumented volume. These are almost entirely due to charged-current (CC) $\nu_{\mu}$ interactions. The second are starting events, where the neutrino vertex is close enough to the instrumented volume that the entire final state is observable, making flavor identification, in principle, possible. All neutrino flavors, and both charged-current (CC) and neutral-current (NC) interactions contribute to this event class.

The transfer tensor for muon events is computed as follows. The flux of neutrino-induced muons arriving at the instrumented volume with energy $E_{\mu}$ from the colatitude angle $\theta$ is given by
\begin{equation}
    \frac{d\Phi_{\mu}(E_{\mu}, \theta)}{dE_{\mu}} = \sum_{x} \int_{0}^{\infty} 
    \frac{d\Phi_{{\nu}_{x}}(E_{{\nu}_{x}}, \theta)}{dE_{{\nu}_{x}}}
    \frac{dP(E_{\mu} | E_{{\nu}_{x}}, \theta)}{dE_{{\mu}}}
    dE_{{\nu}_{x}}
    ,
\end{equation}
where the sum $x$ runs over neutrino flavors, and $P(E_{\mu} | E_{{\nu}_{x}})$ is the probability that a neutrino of flavor $x$ and energy $E_{{\nu}_{x}}$ at the surface of the Earth produces a muon that reaches the instrumented volume with energy less than $E_{\mu}$. This is evaluated by Monte Carlo, propagating neutrinos to the target volume with \texttt{NeutrinoGenerator}~\cite{Gazizov:2004va} and propagating muons produced in charged-current interactions with \texttt{PROPOSAL}~\cite{Koehne:2013gpa}. The final product is a $6 \times 100 \times 40 \times 100$ transfer tensor $T$. Each entry $T_{ijkl}$ gives the probability that a neutrino of type $i$, incident on the Earth with energy $j$ and angle $k$ with respect to the zenith at the detector size, would produce a muon that reaches the instrumented volume with energy $l$. This approach treats the details of muon range straggling correctly~\cite{Lipari:1991ut}, at the expense of fixing the neutrino-nucleon cross-section.

\begin{figure}
    \centering
    \includegraphics[width=0.9\textwidth]{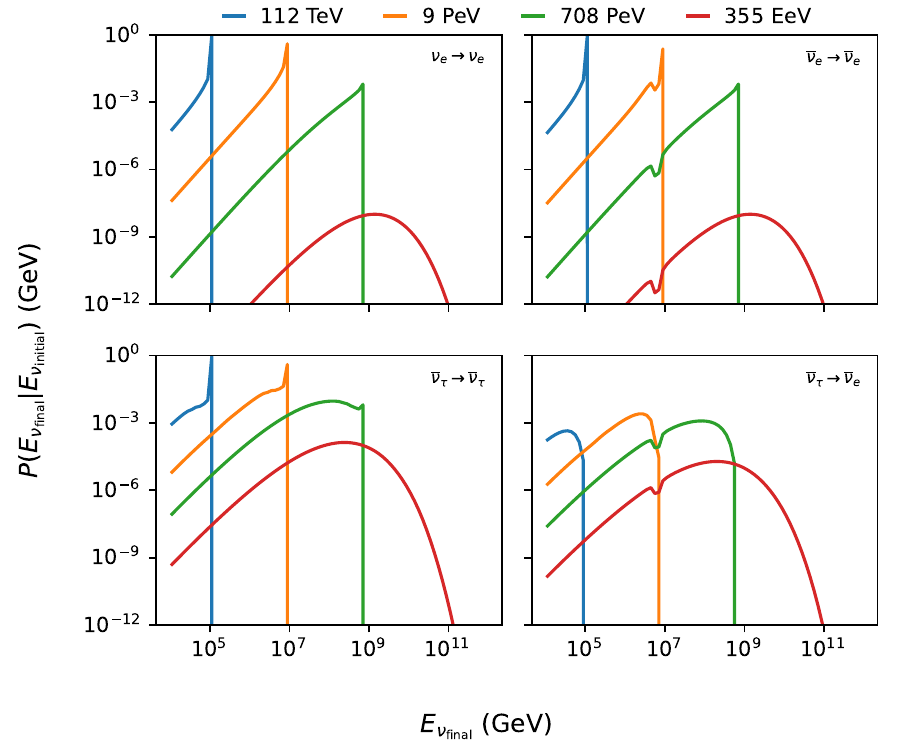}
    \caption{Neutrino flux transfer function evaluated at $\cos\theta = -0.15$. The lines in each panel show the probability of observing a neutrino of the given final flavor as a function of final neutrino energy for different $E_{\nu,{\rm initial}}$. The lower left panel illustrates the ``$\tau$ regeneration'' effect, where decays of $\tau$ leptons produced in CC $\nu_{\tau}$ interactions result in a secondary flux of lower-energy $\nu_{\tau}$.}
    \label{fig:flux-transfer-matrix}
\end{figure}

\begin{figure}
    \centering
    \includegraphics[width=0.9\textwidth]{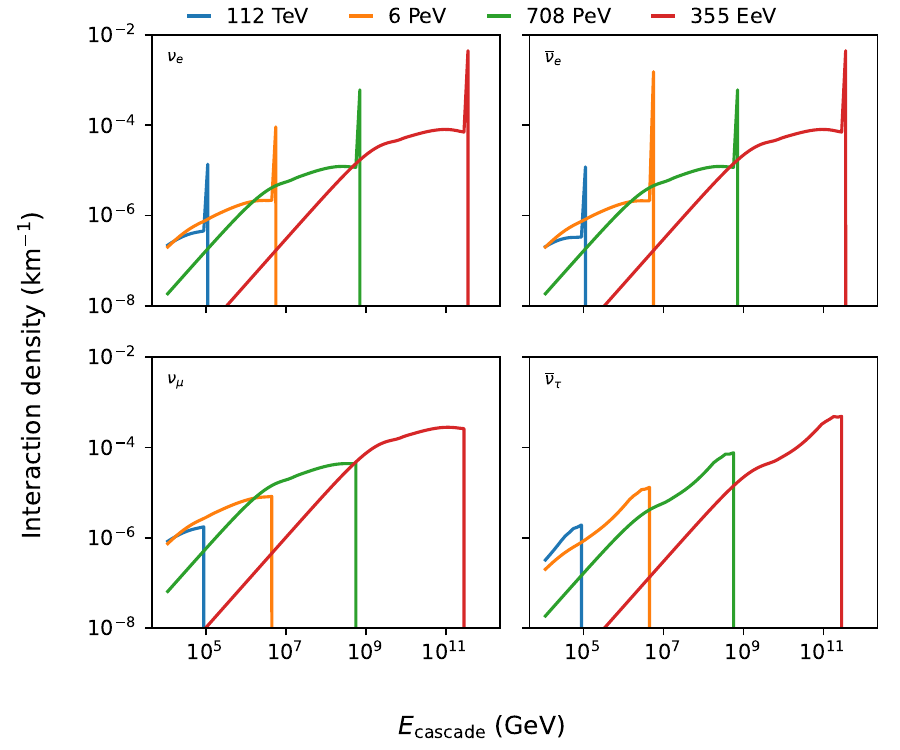}
    \caption{Final state density per surviving neutrino. The lines in each panel show the number of interactions per km of South Pole ice as a function of cascade energy for different $E_{\nu,{\rm final}}$. }
    \label{fig:final-state-density}
\end{figure}

\begin{figure}
    \centering
    \includegraphics[width=0.9\textwidth]{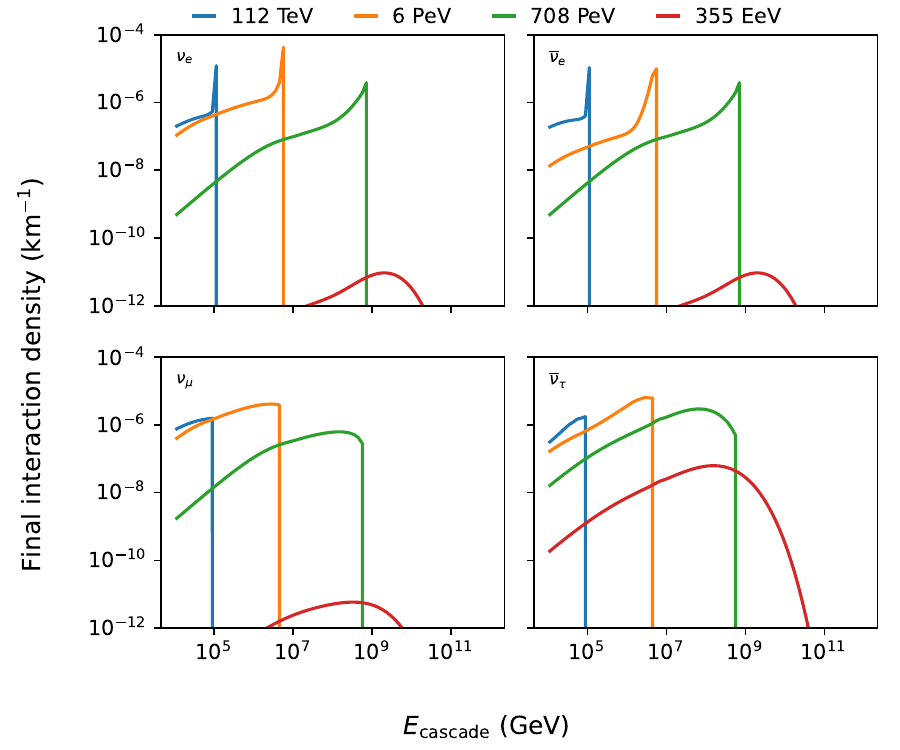}
    \caption{Neutrino-to-cascade transfer function evaluated at $\cos\theta = -0.15$, i.e. final state density per initial neutrino. The lines in each panel show the number of interactions per km of South Pole ice as a function of cascade energy for different $E_{\nu,{\rm initial}}$. }
    \label{fig:final-state-transfer-matrix}
\end{figure}

The transfer tensor for starting events is calculated using a method adapted from Vincent \textit{et al.}~\cite{Vincent:2017svp}. First, we solve the cascade equations for a test flux to obtain a transfer matrix $P$ where $P_{ij} = P(E_{\nu,i} | E_{\nu,j})$, the probability that a neutrino incident on the Earth with energy $E_{\nu,j}$ either reaches the detector with energy $E_{\nu,i}$ or produces a secondary neutrino that reaches the detector with energy $E_{\nu,i}$. For each zenith angle, we produce one matrix for each pair of initial and final neutrino types, resulting in 36 matrices for each zenith angle. Of these, only 10 have nonzero entries: $\nu_{x} \rightarrow \nu_{x}$ and $\overline{\nu}_{x} \rightarrow \overline{\nu}_{x}$ for $x=\{e, \mu, \tau\}$ due to CC losses and NC down-scattering, as well as $\nu_{\tau} \rightarrow \nu_{e}$, $\nu_{\tau} \rightarrow \nu_{\mu}$, $\overline{\nu}_{\tau} \rightarrow \overline{\nu}_{e}$, and $\overline{\nu}_{\tau} \rightarrow \overline{\nu}_{\mu}$ due to secondary $\tau$ decay. We neglect the energy loss of $\tau$ leptons before decay, as well as the contribution of secondary neutrinos from leptonic $W^-$ decays. \autoref{fig:flux-transfer-matrix} shows columns of $P$ for different initial and final neutrino types at a moderate column depth.
Next, we use the differential neutrino-nucleon cross-sections to approximate the production rate of neutrino-induced cascades as a function of neutrino and cascade energy:
\begin{equation}
    \frac{dN}{dE_{c}dl} \approx \frac{d\sigma}{dE_{c}} \rho N_A \Delta E
    ,
\end{equation}
where $\rho$ is the mass density of the detector medium and $\Delta E$ is the width of each energy bin. For CC $\nu_{e}$ and $\overline{\nu}_e$ interactions both the outgoing electron and hadronic cascade contribute; for resonant $\overline{\nu}_e$/$e^-$ scattering, there is an additional contribution from hadronic decays of the resulting $W^-$. As in the transmission calculation, we neglect contributions from leptonic $W^-$ decays. For $\nu_{\mu}$ and $\overline{\nu}_{\mu}$ and NC $\nu_{\tau}$ and $\overline{\nu}_{\mu}$ interactions, only the hadronic cascade contributes. For CC $\nu_{\tau}$ and $\overline{\nu}_{\tau}$ cascades, we consider only cascades from the decay of the secondary $\tau^{\pm}$. The end product is a transfer matrix $I$ for each neutrino type whose entries give the number of cascades produced per unit path length. \autoref{fig:final-state-density} shows columns of $I$ for select neutrino types.
Finally, we form the inner product of $P$ and $I$ to obtain a transfer tensor $T$, where $T_{ijkl}$ gives the average number of cascades that a neutrino of type $i$, incident on the Earth with energy $j$ and angle $k$ with respect to the zenith at the detector site, would produce per unit length of detector medium at cascade energy $l$. \autoref{fig:final-state-transfer-matrix} shows slices of $T$ for select initial neutrino types at a moderate column depth.

This transfer tensor is then multiplied into the final-state effective area described in \autoref{sec:a_eff} to form a neutrino effective area whose first 3 dimensions are the same as those of the transfer tensor (neutrino type, energy, and incidence angle). The transfer tensor is calculated on demand given a parameterization of the neutrino-nucleon cross-section. One such parameterization, created from B-splines fit to cross-sections obtained from \texttt{nusigma} \cite{nusigma} using the CTEQ6 parton distribution \cite{Pumplin:2002vw}, is distributed with \fwn, but other parameterizations may be used in its place.

At the higher energies relevant for radio detection, the transfer matrices described above can still be used as a reasonable proxy. However, since event signatures for the charge current interactions of $\nu_e$ (LPM effect \cite{Migdal:1956tc}) and $\nu_\tau$ ($\tau$ track length) change, and because future proposed detectors are sparsely instrumented, the \fwn{} framework allows for using custom transfer matrices instead. This has the advantage that the deposited shower energy distribution at trigger level for the considered detector can be taken into account directly, at the expense of not using the cross sections and inelasticity distributions implemented within \fwn.

\subsection{Effective Area and Volume}\label{sec:a_eff}

For a neutrino telescope, a few factors control the expected sensitivity to astrophysical phenomena. The most critical of these is the acceptance of the detector to signals and backgrounds. The acceptance is usually estimated as an energy ($E$) and zenith ($\theta$) dependent effective area $A_{\rm{eff}}(E, \theta)$. These acceptances are convolved against the neutrino-to-final state transfer tensor from \autoref{sec:neutrino_physics} to calculate event rates. In this section, we describe the process of calculating and parameterizing the effective area for optical and the effective volume for radio neutrino telescopes. 
The \fwn{} framework implements and manages effective areas through the \texttt{effective\_areas} module. We show details on its use in the examples in \autoref{sec:examples}.

\subsubsection{Optical Neutrino Telescopes}
% (Brian)

To calculate the effective area $A_{\rm{eff}}(E, \theta)$ of the detector, the geometric area $A_{\rm{geo}}(\theta)$ must be multiplied with an energy and zenith dependent efficiency for detecting and selecting events of interest $\eta(E, \theta)$ as in \autoref{equ:optical_eff_area}:

\begin{equation}
    A_{\rm{eff}}(E, \theta) = A_{\rm{geo}}(\theta) \times \eta(E, \theta)
    .
    \label{equ:optical_eff_area}
\end{equation}

For an optical neutrino telescope like IceCube or KM3NeT, the instrumentation is usually deployed in a relatively compact footprint on the scale of a few kilometers. A convex hull can be placed around the geometric extent of the instrument, defining a solid.
Assuming the detector is roughly azimuthally symmetric, the projection of the solid along a given zenith angle $\theta$ defines the geometric area of the detector $A_{\rm{geo}}(\theta)$.  The selection efficiency, $\eta(E, \theta)$, characterizes the triggering efficiency of the detector, as well as the probability with which an event passes a set of analysis criteria. In practice, the selection efficiency is usually estimated using Monte Carlo techniques, where a detector is exposed to an isotropic flux of signal or background events, and triggering simulation and analysis selections are performed. The ratio of the number of events passing these cuts to the number of events thrown defines the selection efficiency. However, simple approximations can also be entered into the framework. 

For the geometric area $A_{\rm{geo}}$, the framework contains a utility module, \texttt{surfaces}, to create and manage the geometric solids, its projections, and azimuthal averages. To construct an object, the user provides a 2D footprint of the detector and an extent, or depth, over which to extrude the footprint to define the bounding volume. 

For the selection efficiency $\eta$, the framework expects the energy and zenith dependent efficiency to be provided as multidimensional B-splines formulated with the photospline package~\cite{Whitehorn:2013nh} and stored as FITS files~\cite{1981A&AS...44..363W}. An example of how to prepare these splines is provided in the code repository~\cite{Example:DataPrep}. Two types of optical selection efficiencies are already supported in the framework. The first is a selection efficiency for \emph{tracks}, which are typically created by through-going muons produced in charged-current $\nu_{\mu}$ interactions, and traditionally drive the sensitivity of searches for the sources of astrophysical neutrinos~\cite{IceCube:2014vjc, IceCube:2019cia, ANTARES:2017dda}. In the framework's  \texttt{effective\_areas} module, efficiency for tracks is handled in the eponymous \texttt{ZenithDependentMuonSelectionEfficiency} class. The second is a selection efficiency for \emph{cascades}, which are created by neutral-current interactions of all neutrino flavors, and charged current interactions of electron and tau neutrinos~\cite{IceCube:2013low, IceCube:2020wum}. These are handled by a \texttt{HECascadeSelectionEfficiency} class.

Combining the geometric area and the selection efficiency creates the effective area. The framework currently has one provided class for tracks: the \texttt{MuonEffectiveArea} class. Users can create their own effective area classes by following the example therein, including creating the associated selection efficiency parameterization. 

\subsubsection{Radio Neutrino Telescopes}

Radio detectors are only sensitive to cascades above $\sim10^{16}$ eV, in contrast to optical neutrino telescopes. Since radio neutrino detectors are sparsely instrumented and the possibility to detect a cascade depends more on the orientation of the Cherenkov cone than on the vertex position, using a geometric area as in \autoref{equ:optical_eff_area} is impractical for radio telescopes. 

Instead it is more convenient to simulate neutrino interactions for each neutrino flavor in bins of neutrino energy $E$ and direction $\theta$ in a simulation volume $V_\mathrm{sim}$. The neutrinos are forced to interact in this simulation volume, and then weighted by their transmission probability through the Earth, or weight $w$.
From these, tabulated neutrino effective volumes $V_{\rm{eff}}(E, \theta)$ can be evaluated by taking the ratio of the sum of weights of events triggering to the number of simulated neutrino interactions $N_\mathrm{sim}$:
\begin{equation}
    V_{\mathrm{eff}}(E, \theta) = 
    \frac{
        \sum w_{\rm{trig}}}
        {N_\mathrm{sim}(E, \theta)}\times V_{\mathrm{sim}}(E, \theta).
\end{equation}

The tables of $V_{\rm{eff}}$ can be passed to the \texttt{radio\_effective\_areas} module, where they are interpolated to match the shape of the transfer tensor $T$ (cf. \autoref{sec:neutrino_physics}) and converted to an effective area through the thin-target approximation:
\begin{equation}
    A_{\mathrm{eff}}(E, \theta) = 
    \frac{
        V_{\mathrm{eff}}(E, \theta)}
        {\mathcal{L}_{\rm{int}}(E)}
\end{equation}
where $\mathcal{L}_{\rm{int}}(E)=m_N/(\rho ~ \sigma_{\nu-N})$ is the interaction length of a neutrino in the Earth, and the cross-section contains the energy dependence.

Instead of using the default transfer tensor $T$ shipped with \fwn{}, the \texttt{radio\_effective\_areas} module allows to include a custom transfer tensor, which -- dependent on neutrino flavor, neutrino energy and direction -- holds the neutrino to shower energy  probability distribution for detected events.

Like with the optical array, there are additional selection, or analysis, cuts.
This is usually imposed by requiring the signal
be strong enough to be identifiable above thermal noise, and fulfill certain reconstruction requirements. Radio neutrino detectors typically rely on single-station reconstruction of the neutrino properties to maximize the chance of seeing ultra-high energy neutrinos. A practical requirement for reconstruction is, for example, detecting signal in at least three antennas. Similarly, observing both the direct radio signal and a secondary signal reflected/refracted in the firn provides a good estimate of vertex distance and thus energy. A third requirement could be to observe a signal in a specific antenna-type, which can help in determining the neutrino direction.

\begin{figure}
    \centering
    \includegraphics[width=0.45\textwidth]{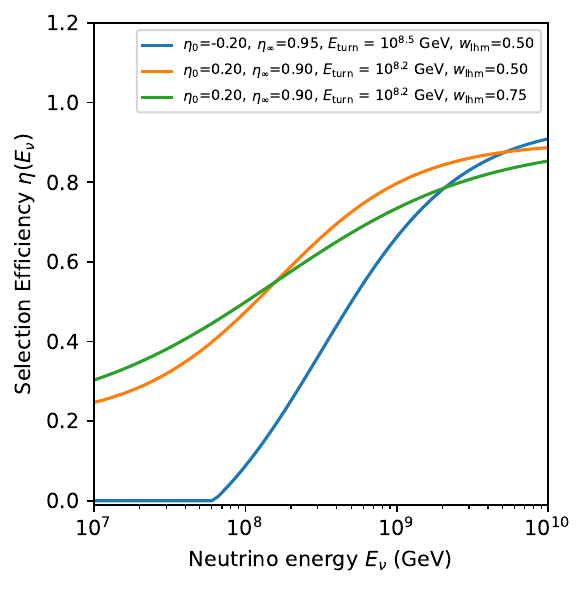}
    \caption{Parameterized efficiencies of the radio array as provided in \fwn{} for different parameter settings as function of energy.}
    \label{fig:analysis_efficiency_radio}
\end{figure}

These additional requirements can be parametrized using an analysis efficiency sigmoid curve in logarithmic energy, as can be seen in \autoref{fig:analysis_efficiency_radio}. This sigmoid/logistic curve shape is common of analysis efficiencies in the radio community~\cite{ARA:2019wcf, Anker:2019rzo}. We have designed the  \texttt{radio\_analysis\_efficiency} module such that the passed parameters correspond to physical quantities, namely the efficiencies in the low and high energy limits ($\eta_{0}$, $\eta_{\infty}$), a turn-on energy $E_\mathrm{turn}$ where 50\% of the maximum is reached, and the logarithmic width\footnote{More precisely, the region covering 25\% to 75\% of the transition from $\eta_{0}$ to $\eta_{\infty}$} of the turn-on $w_\mathrm{lhw}$,  
\begin{equation}\label{eq:radio_analysis_efficiency}
      \eta(E) = \frac{\eta_{0} - \eta_\infty}{1+\exp\left(w_\mathrm{lhm}/\log(3) * \log_{10}(E / E_\mathrm{turn})\right)} + \eta_\infty.
\end{equation}

The analysis efficiency may be chosen such that it also effectively reduces the contribution of the non-physics backgrounds, which can originate from high-wind periods, human activity near the detector or thermal noise surpassing the trigger threshold. 

By choosing the different parameterization for \autoref{eq:radio_analysis_efficiency} we can use \fwn{} to evaluate how optimistic or conservative assumptions of analysis efficiency impact the instrument's sensitivity.
In particular, if the analysis efficiency behavior with respect to zenith is known, a user can define a more sophisticated parameterization taking the zenith dependency into account.

\subsection{Angular and Energy Resolution}
\label{sec:detector_resolution}

In addition to the effective area and selection efficiencies, the detector sensitivity depends on performance metrics such as the angular and energy resolution. In this section we discuss how the angular and energy resolutions are parameterized and utilized in the framework. The angular resolutions are handled in the \texttt{angular\_resolution} module, while the the energy resolutions are managed by the \texttt{energy\_resolution} module.

\subsubsection{Optical Neutrino Telescopes}

\begin{figure}
    \centering
    \includegraphics[width=0.7\textwidth]{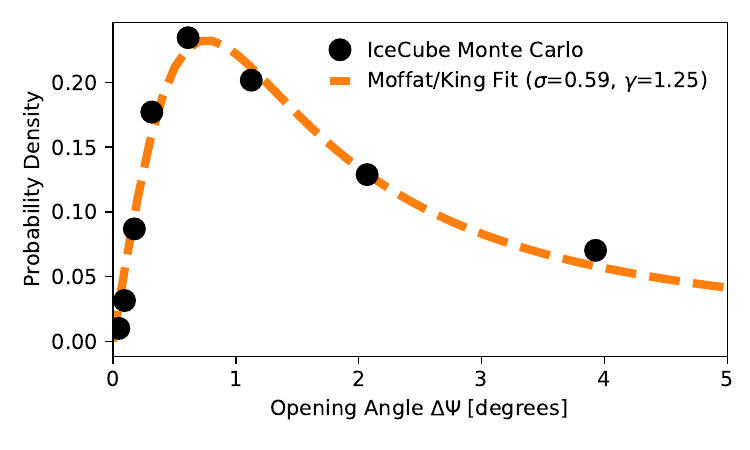}
    \caption{An example of a point spread function from IceCube (black points), and the fitted Moffat/King parameterization utilized by the framework (dashed line). The Monte Carlo points are taken from the public data release~\cite{IceCube:2021xar} associated with IceCube's latest search for point sources using ten years of data~\cite{IceCube:2019cia}, specifically the distribution for $5 \leq \log_{10}(E_{\nu}) \leq 5.5$ and declinations $\delta < -10$ (corresponding to the Northern sky).}
    \label{fig:optical_psf_and_fit}
\end{figure}

Like the effective areas and selection efficiencies, the framework expects the optical angular resolution to be provided as energy and zenith dependent B-splines in photospline~\cite{Whitehorn:2013nh}. Again under the assumption of azimuthal symmetry, the angular resolution, described through the \textit{point spread function} (PSF), reduces to an energy and zenith dependent quantity. The PSF is defined by the distribution of opening angles $\Delta \Psi$, where $\Delta \Psi$ is the angle between reconstructed event direction and the true direction of the neutrino. For the optical array, the framework currently supports parameterizing the PSF using the \emph{Moffat/King} function, as is done in e.g.\ the \textit{Fermi} gamma ray telescope~\cite{1969A&A.....3..455M, 1962AJ.....67..471K, FermiPSF}. For a given energy and zenith bin, the Moffat/King function takes the form of \autoref{equ:moffat_king}:
\begin{equation}
    K(\Delta \Psi, \sigma, \gamma) = \frac{1}{2 \pi \sigma^2} \bigg( 1 - \frac{1}{\gamma} \bigg) \bigg[ 1 + \frac{1}{2\gamma} \frac{(\Delta \Psi)^2}{\sigma^2} \bigg]^{-\gamma}
    .
    \label{equ:moffat_king}
\end{equation}
The function has two parameters, $\sigma$ and $\gamma$. Qualitatively, $\sigma$ describes the overall scale of the PSF (it is analogous to the width of a Gaussian distribution), while $\gamma$ controls the slope of the tail of the distribution. An example of a PSF and the fitted Moffat/King function for a set of public IceCube data is shown in \autoref{fig:optical_psf_and_fit}.
We observe that this parameterization works well for analyses focusing on tracks, which typically have good, degree-scale median angular errors, but a user could implement other parameterizations as necessary. In practice, to calculate the PSF, one applies an analysis event selection to a Monte Carlo signal or background simulation, and builds the multidimensional distribution of $\Delta \Psi$; this can then be fit with a multidimensional B-spline, and saved to a FITS file. An example of building the multidimensional spline is provided in the code repository~\cite{Example:DataPrep}.

Construction of the energy resolution parameterization proceeds analogously, where a user applies cuts, and then builds distributions of reconstructed energy as a function of true energy. The framework particularly works with \textit{final state} energy, e.g. energy of the muon at the detector boundary, or energy deposited in a cascade. The transfer function between this observable final state and the true flux of atmospheric or astrophysical neutrinos is discussed in \autoref{sec:neutrino_physics}. 
The framework currently supports a generic \texttt{EnergySmearingMatrix} class, where for a true final state energy bin, a user provides a bias and width for the reconstructed energy. We observe that for optical neutrino telescopes, e.g. IceCube~\cite{IceCube:2013dkx} and ANTARES~\cite{ANTARES:2017wlx}, a parameterization in energy alone is sufficient, as the energy resolution tends not to have strong zenith or azimuth dependence. Because this parameterization is comparatively simple, and requires only a univariate spline, there are minimal memory and computational savings to storing splines instead of the binned resolution directly, and so currently the framework uses the latter. The default class expects the bias and width parameters to be supplied as compressed archived files, specifically Numpy \texttt{npz} files, and a spline is built when an instance of the class is created. One implementation is already provided in the  \texttt{MuonEnergyResolution} sub-class, which focuses on tracks. A user could build their own classes with more sophisticated parameterizations.

\subsubsection{Radio Neutrino Telescopes}

\begin{figure}
    \centering
    \includegraphics[width=\columnwidth]{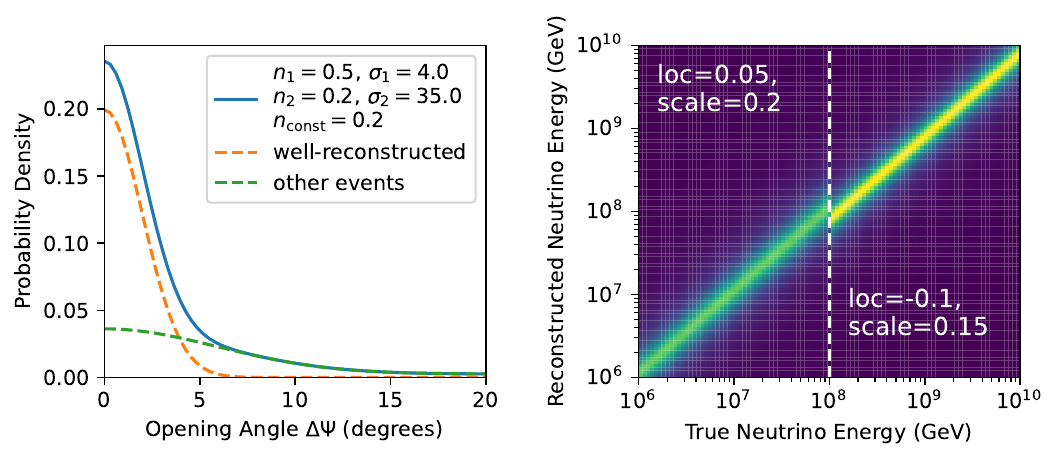}
    \caption{Angular and energy response for the radio detector as currently used in \fwn. The left figure shows the probability as function of opening angle between true arrival direction and reconstruction. It is a combination of well-reconstructed and other neutrinos. The rights side shows two parameterization of the reconstructed energy as function of true energy.}
    \label{fig:radio_energy_angular_resolution}
\end{figure}

Reconstruction methods for radio neutrino detectors are under development but are currently less mature than for optical telescopes \cite{Aguilar:2021uzt,RNO-G:2021zfm}. In particular, detailed large sets of Monte Carlo simulations after reconstruction are often not present. Therefore, more general smearing functions are currently provided in the \texttt{radio\_responses} module for the angular error and energy smearing. The currently implemented functions for the radio energy and angular resolution are shown in \autoref{fig:radio_energy_angular_resolution}.

As opposed to optical telescopes, to maximize effective volume, radio detectors are often designed to reconstruct a neutrino event with only a single station. Whether a neutrino event is detected or not depends primarily on the vertex distance \cite{Aguilar:2021uzt} and is therefore, to first order, independent of the neutrino energy. The \texttt{RadioEnergyResolution} class uses a Cauchy PDF to account for the resolution on the shower energy, which compared to a Gaussian has larger tails, and therefore matches the distribution of reconstructed events better. 

Accurate pointing for the neutrino direction is typically only possible for events for which the shower vertex position and the signal polarization can be determined. 
The \texttt{RadioPointSpreadFunction} describes the angular resolution in terms of $\Delta\Psi$ using a double-Gaussian PDF with a constant term. The parameters are the widths of the Gaussians $\mathcal{G}$, ($\sigma_1$, $\sigma_2$), and the fraction of events in each component ($n_1$, $n_2$, $n_\mathrm{const}$),
\begin{equation}
    R(\Delta\Psi, (n_1, \sigma_1), (n_2, \sigma_2), n_\mathrm{const}) =
    \frac{n_1\cdot\mathcal{G}(\sigma_1) + n_2\cdot\mathcal{G}(\sigma_2) + n_\mathrm{const}}{n_1+n_2+n_\mathrm{const}} 
\end{equation}
While the first Gaussian accounts for events that can be accurately reconstructed, the remaining events are absorbed in the second Gaussian and constant terms. With future and more detailed analyses based on larger Monte Carlo sets, a more realistic description of the angular resolution will give some guidance on how to incorporate the dependence on the true zenith direction into the angular resolution. 

\subsection{Background}\label{sec:backgrounds}

When estimating sensitivities, we must account for backgrounds that may mimic the signal by contributing to the event rate in a region of interest. Which categories of background are important depend both on the kind of detector, the event selection, and the target science case. For a measurement of the diffuse astrophysical neutrino flux with an optical detector, for example, the primary backgrounds are penetrating atmospheric muons and, to a lesser degree, atmospheric neutrinos. In a search for point sources of neutrinos, on the other hand, the overall flux of high-energy astrophysical neutrinos is itself a background. If this were not accounted for, a single upward-going muon of sufficiently high energy could be considered evidence of point emission.

Backgrounds can be accounted for either by adding their contributions to the event rate or ignoring regions where they are expected to contribute. The framework takes both approaches in different cases. For all detectors and science cases, the flux of atmospheric neutrinos (as parameterized with \texttt{nuflux} \cite{nuflux}) is added as a background contribution, using the same effective area as for astrophysical neutrinos. The atmospheric neutrino flux can optionally be reduced in the downward-going region to account for the fact that atmospheric neutrinos can be vetoed by the detection of accompanying muons from the same extensive air shower \cite{Schonert:2008is}, following the calculation of Gaisser \textit{et al.}~\cite{Gaisser:2014bja}.

For the detection of incoming muons with an optical detector, an approximate flux of single, penetrating muons from extensive air showers is added as a background contribution in the downward-going region. This can also be reduced to account for an ability to veto penetrating muons by detecting the electromagnetic component of the extensive air shower at the surface over a limited solid angle, e.g. with IceCube and IceTop \cite{IceCube:2016tpw}. For starting events with an optical detector, an energy threshold is applied such that the penetrating muon background is negligible compared to the rate of high-energy astrophysical neutrinos, mimicking the event selection of e.g. IceCube's ``High Energy Starting Events" sample~\cite{IceCube:2013low}. 

A potential irreducible background to neutrino detection with under-ice radio antennas is possible from extensive air showers penetrating the ice or from ultra-high-energy atmospheric muons producing detectable showers deep in the ice. The rate at which these events occur is subject to large uncertainties and actively debated, but roughly follows the zenith angle distribution of the neutrino signal \cite{Garcia-Fernandez:2020dhb}. If specified, the framework expects the assumed air-shower background rate as a two-dimensional grid binned in in-ice shower energy and zenith angle ($\cos\theta$). We assume that all other reducible backgrounds, such as triggers induced by noise fluctuations, wind \cite{Mikhailova:2021ccy}, or human activity, are effectively suppressed by the applied selection efficiency.

\section{From detector parameters to physics performance}
With a full description of the detector and the expected backgrounds, the physics performance can be evaluated. We first describe methods and approach before discussing the general set-up of the examples \autoref{sec:examples}, which are each described in \autoref{appendix_examples}.

\subsection{Framework Overview}

Once the response of a detector is fully described in the response tensor $D_{f,E_\mathrm{\nu},\cos(\theta_\nu),E_\mathrm{rec},\Delta\Psi}$, it can be stored in the \fwn{}\texttt{.factory}. In addition to the $D_{f,E_\mathrm{\nu},\cos(\theta_\nu),E_\mathrm{rec},\Delta\Psi}$, the factory allows the user to pass a distribution of non-neutrino backgrounds as described in \autoref{sec:backgrounds} for each specified detector. 
The \texttt{factory} is used to collect the responses of multiple instruments (e.g.\ optical and radio) or disjoint analysis samples of one instrument (e.g.\ cascades and (un)shadowed tracks for an optical detector). The detectors stored in the \texttt{factory} can then later be retrieved for chosen livetimes. This allows the user to retrieve several stored effective area tensors and join them to produce combined sensitivity estimates. In addition, the same (set of) detector(s) can be retrieved for a series of livetimes in order to evaluate the sensitivity evolution with longer operation. Examples of retrieving a combination of detectors from the \texttt{factory} are given in \autoref{sec:examples}. As can be noted throughout the previous sections, the optical and radio component of the detector are conceptually different, so the radio components of the detector response are often isolated into radio-specific modules, like the \texttt{radio\_aeff\_generation}. Parameters for the radio related modules can be defined in corresponding YAML files, and then manipulated by the framework; this is further discussed and demonstrated in \autoref{sec:radio_ex}.

Neutrino fluxes are multiplied with the effective area tensors and their requested livetimes to obtain the expected event rates. Frequently used neutrino flux models for diffuse (atmospheric and astrophysical) neutrino fluxes and point source flux predictions are shipped together with the framework in the \texttt{diffuse} and \texttt{pointsource} modules, respectively.

The user may define source and background fluxes contributing to a particular analysis, for example, a diffuse astrophysical neutrino spectrum with a background of atmospheric neutrinos; or a point source spectrum with the sum of diffuse astrophysical and atmospheric neutrinos as backgrounds. The atmospheric muon background described in \autoref{sec:backgrounds} adds to the neutrino background. The event rates for signal and background can be used in an \emph{Asimov} approach~\cite{Cowan:2010js} to estimate the sensitivities for the set of detectors and livetimes obtained from the factory; see the following section for a more thorough discussion.

\subsection{Method to estimate sensitivities}

In estimating the sensitivity of a detector to astrophysical phenomena, a few quantities are typically used. For example, it is common for a detector to quote the \textit{discovery potential} for a flux of neutrinos from a source or a class of sources. The discovery potential is usually defined as the minimum value required of a parameter of the astrophysical flux, e.g.\ the normalization, such that an experimental search would exclude the parameter as being zero at the $5\sigma$ level. 
We can also define a median 90\% confidence level (CL) upper limit on a parameter, which is sometimes referred to as the sensitivity. It should be noted that discovery potential and sensitivity are distinct; discovery potential quantifies the ability of an experiment to discover a real signal, while sensitivity the ability to exclude one.  

Experimental searches and measurements are usually implemented as Poisson maximum-likelihood problems, where a likelihood function is constructed of terms that model the backgrounds (such as the atmospheric neutrino flux) and terms that model a signal (such as the spectral index of the neutrino flux)
The framework contains a module, \texttt{multillh}, and a contained class, \texttt{LLHEval}, for facilitating these log-likelihood type analyses. It contains utilities for specifying fixed and free (nuisance) parameters, and evaluating, fitting, and profiling over the likelihood space. The framework uses the SciPy limited-memory Broyden-Fletcher-Goldfarb-Shanno (L-BFGS-B) routine for the fitting~\cite{2020SciPy-NMeth}. Several examples of how to use the likelihood tool to estimate performance for both diffuse and point-source type analyses are provided in the examples in \autoref{sec:examples}.

For example, the Poisson likelihood $\mathcal{L}$ in the search and characterization of a diffuse flux of neutrinos takes the form of:
\begin{equation}
    \mathcal{L}(n|\mu(\vec{\theta}, \vec{\xi}))= \prod_{i=1}^{N} \frac{(\mu_i)^{n_i}}{n_i!}e^{-\mu_i}
    .
    \label{equ:likelihood_diffuse}
\end{equation}
In this formulation, $n$ is the measured number of events in some bin $i$, and $\mu$ is the expected number of events given parameters of the astrophysical flux $\vec{\theta}$ and nuissance parameters of the backgrounds (e.g. the atmospheric flux) $\vec{\xi}$. The best fit signal parameters $\hat{\vec{\theta}}$ can be found by maximizing the likelihood. The ratio of the log of the likelihood at the best fit point to that of the likelihood under the null hypothesis (typically $\vec{\theta}=0$, meaning no astrophysical component is present) defines the test statistics $\lambda = 2 \log \mathcal{L}(\vec{\theta}=\hat{\vec{\theta}})/\log \mathcal{L}(\vec{\theta}=0)$.

To evaluate the significance of the observation 
one typically performs pseudo-experiments (PEs) to establish the expected distribution of $\lambda$ under the null hypothesis. The observed test-statistic can then be compared to the background-only distribution to determine the significance of an observation. A PE is usually formed by drawing realizations of the data Poisson-distributed around the mean event rate in an observable bin, and then performing the analysis on the pseudo-data. Through \emph{signal-injection} or artificially adding simulated signal events to the data sample, one also derives the sensitivity and discovery potential. For example, one can use this process to identify the largest possible flux that would still not be detected at the 90\% CL. By running many PEs, one builds a distribution of 90\% CLs, and the median of this distribution becomes the quoted sensitivity.

Unfortunately, the process of generating many pseudo-experiments is too computationally expensive for purposes of rapidly evaluating sensitivities in the framework. To make the estimates computationally feasible, the framework makes two simplifying assumptions. First, the framework builds an \emph{Asimov} dataset, which replaces the observed event rates by the exact mean~\cite{Cowan:2010js}. Second, the framework assumes the distribution of the null test-statistic is $\chi^2$ distributed as predicted by Wilk's Theorem~\cite{Wilks:1938dza}. These assumptions are justifiable as the test-statistic derived from the Asimov set is asymptotically equivalent to the median test-statistic derived from Poisson-sampled realizations~\cite{Cowan:2010js, Wald1943TestsOS}, and the event rate in the bins providing most of the test-statistic is large ($\gtrsim 10$).

\subsection{Application examples and fictive detector}
\label{sec:examples}
Since this framework is intended to be of practical use, it is shipped with four examples of obtaining figures and quantities that can be used to evaluate the performance of a neutrino detector. Each example is contained in an \textit{iPython} notebook obtainable in the GitHub repository~\cite{Toise_notebooks}. They are also described in \autoref{appendix_examples}.

The numerical characteristics provided in these examples are those of a fictive detector. While this framework can describe the detector performance of any existing detector, the default values in the examples should not be considered as description of an actual detector. We look forward to a data release from the IceCube-Gen2 collaboration, as well as other collaborations, providing parameterizations for specific instruments.

In order to quantify the sensitivity of a given detector configuration with detail, the effective areas and instrument responses (cf.\ \autoref{sec:a_eff} and \autoref{sec:detector_resolution}) need to be quantified. 
As a starting point and example, the software package contains simplified detector response curves for an under-ice optical array and a radio station, which is instructive to discuss to understand various aspects of detector performance that are required instead of simply using effective areas.  

\begin{figure}
    \centering
    \includegraphics[width=0.9\textwidth]{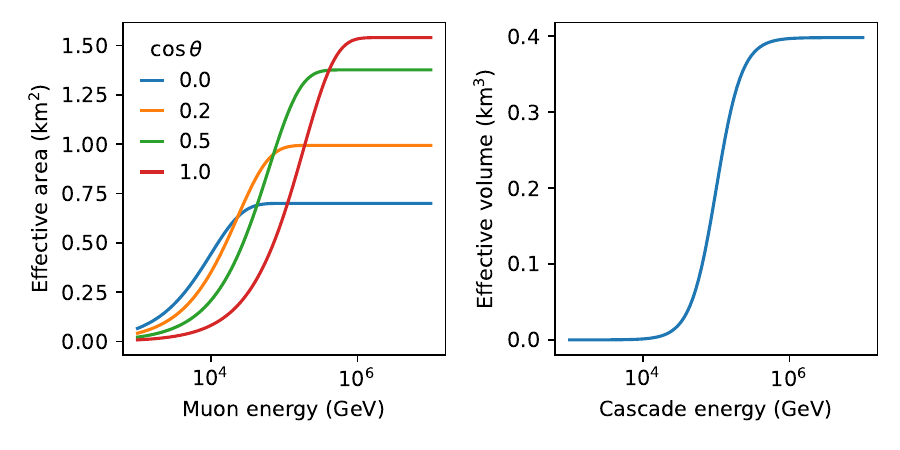}
    \caption{Effective area and volume of the fictive optical detector. The energy threshold for incoming muons increases in vertical directions to account for the decreased chance of a track passing close to an instrumented string. The energy threshold for area starting events is independent of zenith angle.}
    \label{fig:potemkin-optical-aeff}
\end{figure}

\begin{figure}
    \centering
    \includegraphics[width=0.9\textwidth]{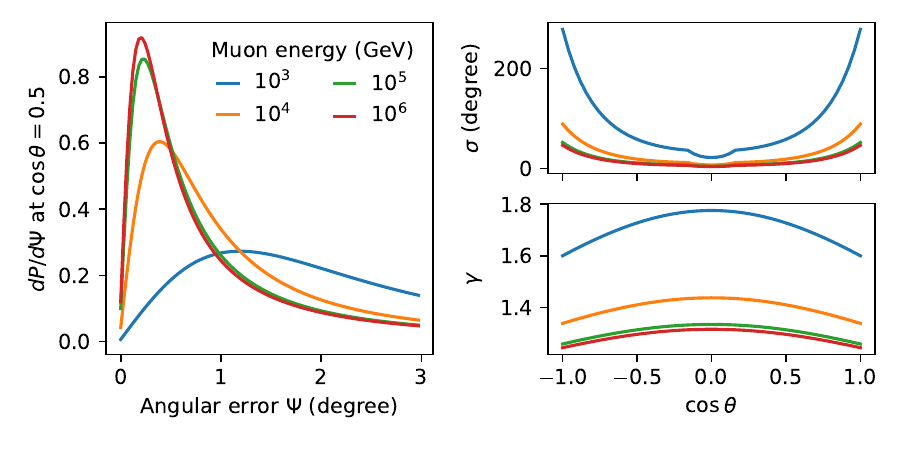}
    \caption{Muon track angular reconstruction performance of the fictive optical detector. The point-spread function becomes narrower (small $\sigma$) and more Gaussian (small $\gamma$) with increasing muon energy. The best angular resolution is achieved for horizontal directions ($\cos\theta=0$); the worst for vertical directions $\cos\theta=\pm 1$.}
    \label{fig:potemkin-optical-muon-psf}
\end{figure}

\begin{figure}
    \centering
    \includegraphics[width=0.9\textwidth]{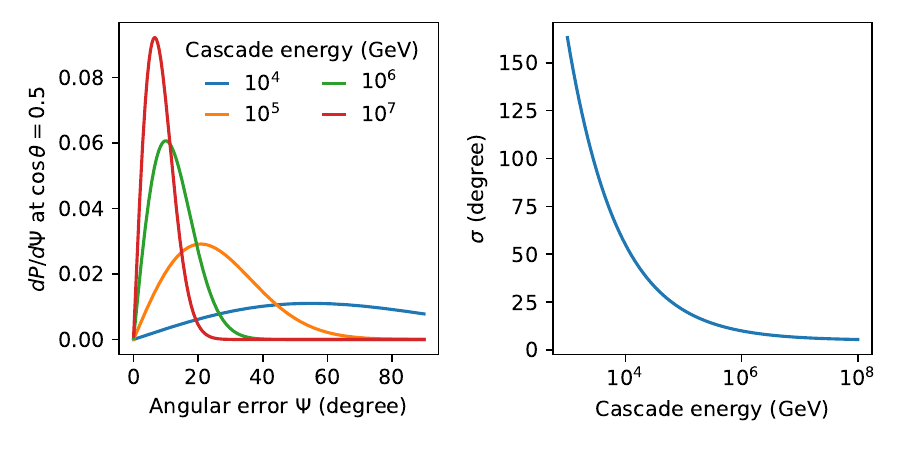}
    \caption{Starting event angular reconstruction performance of the fictive optical detector. The point-spread function becomes narrower (small $\sigma$) with increasing energy, and is independent of zenith angle.}
    \label{fig:potemkin-optical-cascade-psf}
\end{figure}

\begin{figure}
    \centering
    \includegraphics[width=0.45\textwidth]{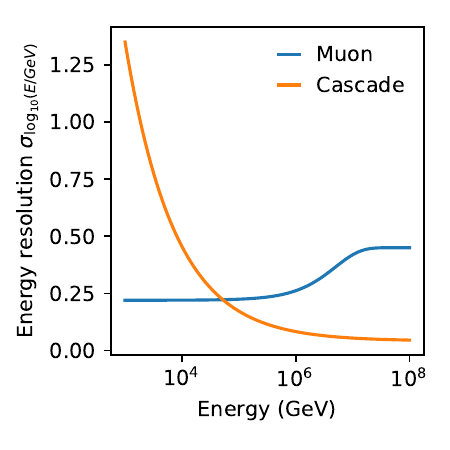}
    \caption{Energy resolution of the fictive optical detector. The (logarithmic) energy resolution for muons is nearly constant at low energies and increases above 100 TeV, modeling the performance of an energy estimator that is primarily sensitive to the mean energy loss rate, but does not account for stochasticity. The energy resolution for starting events is inversely proportional to the square root of energy, modeling an estimator whose performance is dominated by the statistics of detected photons. Energy resolutions are assumed to be independent of zenith angle.}
    \label{fig:potemkin-optical-eres}
\end{figure}

\begin{figure}
    \centering
    \includegraphics[width=0.45\textwidth]{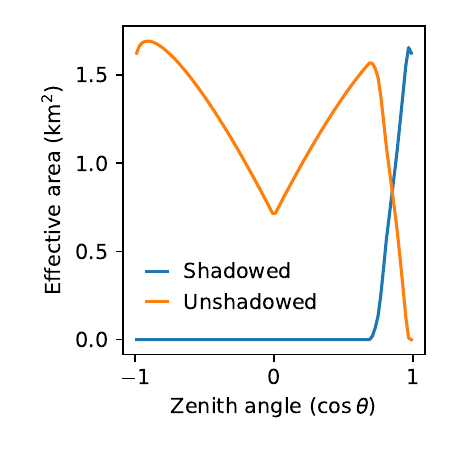}
    \caption{Surface veto coverage of the fictive optical detector. Trajectories from zenith angles less than 47 degrees may pass through both through the footprint of the surface detector and the deep instrumented volume, creating a ``shadowed'' area with reduced penetrating muon background in the throughgoing-track channel. In the remaining, ``unshadowed'' area, the penetrating muon background is unaffected.}
    \label{fig:potemkin-optical-veto-coverage}
\end{figure}

The fictive optical array is based on a cylindrical instrumented volume 500 m in height and 700 m in radius. Its performance metrics have features that could be expected from a detector that is wider than it is tall, and instrumented more densely in the vertical direction than in the horizontal. The effective area for muons shown in \figureautorefname~\ref{fig:potemkin-optical-aeff} is largest from vertical directions, but has a higher energy threshold, accounting for the smaller chance of a muon passing close to a sensor. The angular resolution for muon tracks shown in \figureautorefname~\ref{fig:potemkin-optical-muon-psf} degrades sharply from vertical directions for the same reason. The effective volume and angular resolution for starting events shown in \figureautorefname~\ref{fig:potemkin-optical-aeff} and \ref{fig:potemkin-optical-cascade-psf} are independent of zenith angle, as they depend only on the instrumentation density near the neutrino interaction vertex. Energy resolution (\figureautorefname~\ref{fig:potemkin-optical-eres}) depends mostly on the characteristics of typical energy estimators, and so is assumed to be entirely a function of muon or cascade energy. The fictive detector also includes a surface component with an area of 5~km$^2$ that can detect extensive air showers that pass through the under-ice instrumented volume at zenith angles of less than 47 degrees. \figureautorefname~\ref{fig:potemkin-optical-veto-coverage} shows the area of the under-ice detector that is shadowed by the surface detector.

As a fictive radio detector, we consider a 30 station radio array, where each station is a single dipole at a depth of $\SI{100}{m}$ antenna. This was simulated using the NuRadioMC framework \cite{Glaser:2019cws}. This is a reasonable proxy for a trigger at this depth, when not including overlap between stations, i.e.\ full array effects.

\section{Summary and Outlook}
In this paper, we have presented the \fwn{} framework, which is used to calculate the physics performance of neutrino detectors based on parameterized performance characteristics. The framework code and the related examples are available on GitHub~\cite{toise}.
Since the framework relies on parameterizations of detector performance, using \fwn{} allows for a fast comparison of different detector designs, which is particularly helpful in early stages of experiment planning and design. 
Since next generation instruments like IceCube-Gen2 are foreseen as combination of optical and radio technologies, the \fwn{} framework has been extended to combine different detectors and evaluate their performance together. This feature may also be useful in the global context of exploring the usefulness of joint analyses between different experiments.

The \fwn{} framework has been used by the IceCube-Gen2 collaboration to describe the science capabilities of the detector \cite{IceCube-Gen2:2020qha}.
It will continue to be used for the Technical Design Report (TDR), which is currently in preparation. 
We encourage IceCube-Gen2 to issue a data release of the detector quantities used by \fwn{} in parallel to the TDR.
This will allow the community to evaluate the envisioned instrument performance for different science scenarios. 
Because the framework is general, we also encourage other projects under development, such as P-ONE~\cite{Resconi:2021ezb} or RNO-G~\cite{RNO-G:2020rmc}, to consider releasing detector performance parameterizations in a way that can be utilized by \fwn.

\section{Acknowledgments}
\fwn{} was first developed in the context of IceCube and with an eye towards IceCube-Gen2. We thank our collaborators for adopting this framework and making suggestions for its improvement. We particularly thank Tianlu Yuan for first introducing radio detectors in \fwn{} and Daniel García-Fernández for adding NuRadioMC support.

B.~A.~Clark thanks the National Science Foundation for support through the Astronomy and Astrophysics Postdoctoral Fellowship under Award 1903885. B.~A.~Clark and R.~Halliday thank the Institute for Cyber-Enabled Research at Michigan State University for computing resources. S.~Hallmann and A.~Nelles are supported by the Helmholtz Association (Initiative and Networking Fund, W2/W3 Program).
    
We are thankful to the broader scientific computing community for developing and maintaining the software on which  \fwn{} depends, including  Matplotlib~\cite{Hunter:2007}, Numpy~\cite{derWalt:2011}, SciPy~\cite{Jones:2001}, \textit{healpy}/\textit{HEALPix}~\cite{Zonca2019}, and the conda-forge packaging infrastructure~\cite{conda_forge_community_2015_4774217}.

\bibliographystyle{JHEP}
\bibliography{main}

\appendix

\section{Application Examples}
\label{appendix_examples}
We describe four examples that are provided in \fwn{} that best illustrate the capabilities of the framework. Please see the notebooks in the framework for further details~\cite{Toise_notebooks}. 
\subsection{Ultra-High Energy Differential Sensitivity and Upper Limits}
\label{subsec:uhe_ex}
This first example describes finding a differential sensitivity for the fictive detector configuration.  This first tutorial notebook walks through the basics of \fwn, starting with generating effective areas from preset detector configurations. 

The easiest way to use pre-made instrument responses is through the \texttt{toise.factory} module. This module keeps track of known instrument configurations, and creates effective areas for them on demand. 
As discussed in \autoref{sec:description_intro}, the last dimension of the \fwn{} effective area tensor is the direction resolution binning.
For a diffuse analysis, where we are interested in events across the whole sky, regardless of their angular resolution quality, we we will not constrain this part of the tensor for now, and so choose only a single bin in \texttt{psi} which spans the entire sky from zenith angles of $0$ to $\pi$:

\begin{minted}[
fontsize=\footnotesize,
]{python}
import numpy as np
from toise import factory
factory.set_kwargs(psi_bins={k: [0, np.pi] for k in ('tracks', 'cascades', 'radio')})
\end{minted}
From this we can explicitly call the factory to construct the effective areas for various event classes:
\begin{minted}[
fontsize=\footnotesize,
]{python}
aeffs = factory.get('Fictive-Optical')
\end{minted}
This returns a dictionary whose keys are the event classes. Currently there are:
\begin{itemize}
\item \texttt{shadowed\_tracks}: tracks entering the detector from the outside that pass through the footprint of the surface veto. These have a reduced penetrating muon background.
\item \texttt{unshadowed\_tracks}: all remaining throughgoing tracks.
\item \texttt{cascades}: neutrino interactions inside the fiducial volume (1/2 string spacing inside the outer strings) where the outgoing lepton is a) not a muon, and b) not a tau with a decay length $>$ 300 m.
\end{itemize}
Behind each of these keys is a pair of effective areas, one for neutrinos, and one for penetrating atmospheric muons. Cascades are a special case, as we currently assume that the outer-layer veto completely removes penetrating muons (see \autoref{sec:backgrounds}).

To predict the sensitivity of a given detector configuration (exposure, i.e. effective area and livetime) to a neutrino emission scenario, we have to calculate event rates, which in turn requires a flux model. There are a number of flux models available in \texttt{toise.diffuse} and \texttt{toise.pointsource}. The diffuse versions predict event rates integrated over zenith angle bins, while event rates from point source fluxes are divided into rings centered on the putative source position.

A real analysis will use multiple detection channels in different detectors (e.g.\ $N$ years of optical array plus $M$ years of radio array), each of which is represented by a different effective area. We can create event rate predictions for such a collection of effective areas by writing a factory function and 
using \texttt{factory.component\_bundle()} to construct components that predict event rates in all detectors and detection channels for the given combination of livetimes (here, 15 years).

\begin{minted}[
fontsize=\footnotesize,
]{python}
from toise import diffuse, multillh, surface_veto

def make_components(aeffs):
    aeff, muon_aeff = aeffs
    
    energy_threshold = np.inf
    atmo = diffuse.AtmosphericNu.conventional(aeff, 1, hard_veto_threshold=energy_threshold)
    atmo.prior = lambda v, **kwargs: -(v-1)**2/(2*0.1**2)
    prompt = diffuse.AtmosphericNu.prompt(aeff, 1, hard_veto_threshold=energy_threshold)
    prompt.min = 0.5
    prompt.max = 3.
    astro = diffuse.DiffuseAstro(aeff, 1)
    astro.seed = 2.
    
    if muon_aeff is None:
        muon = diffuse.NullComponent(aeff)
    else:
        muon = surface_veto.MuonBundleBackground(muon_aeff, 1)
    
    return dict(atmo=atmo, prompt=prompt, astro=astro, muon=muon)

    return components
    
bundle = factory.component_bundle({'Fictive-Optical': 15}, make_components)
components = bundle.get_components()
\end{minted}

We can use these to construct a likelihood and predict sensitivities and discovery potentials. As an illustration, we calculate the sensitivity (median 90\% upper limit assuming there is no actual signal) for the normalization of an E$^{-2}$ isotropic equal-flavor flux. To calculate sensitivity, we create pseudodata where there is no contribution at all from the E$^{-2}$ component and calculate the test-statistics (TS) between the best fit (astrophysical normalization is zero) and a given normalization, and finding the point where this crosses $\sim 2.705$ (90\% CL for 1 degree of freedom):

\begin{minted}[
fontsize=\footnotesize,
]{python}

llh = multillh.asimov_llh(bundle.get_components(), astro=0)

from scipy.optimize import bisect
from scipy import stats
# test statistic between astro = f and astro = 0
ts = lambda f: -2*(llh.llh(**llh.fit(astro=f)) - llh.llh(**llh.fit(astro=0)))
# fit for \Delta TS = 2.705 (90% CL for 1 degree of freedom)
critical_ts = stats.chi2(1).ppf(0.9)
limit = bisect(lambda f: ts(f)-critical_ts, 0, 1)
print(limit)
> 0.0079
\end{minted}

This result indicates that if the flux of astrophysical neutrinos were actually zero, 15 years of observing with the fictive detector would constrain that flux of astrophysical neutrinos to $E^2\Phi < 0.0079 \times 10^{-8} \,\, {\rm GeV \, cm^{-2} \, sr^{-1} \, s^{-1}}$ per neutrino flavor. This is a median upper limit; depending on the fluctuations in the number of observed atmospheric background events, the actual flux limit could be higher or lower, each with probability 0.5.

This component-construction and limit-setting procedure is automated for a few common cases in the \texttt{figures\_of\_merit} module, including for
an upper limit (sensitivity), discovery potential, or a Feldman-Cousins upper limit, which is more useful in the ``no-backgrounds" case typical of searches for neutrinos at ultra-high energies. These can be either integrated over the entire energy range of the signal, or divided into energy bins. For example, we make a figure-of-merit object for testing sensitivity to the Ahlers \& Halzen cosmogenic neutrino flux \cite{Ahlers:2012rz} or we can calculate a model-dependent upper limit using the full energy range:

\begin{minted}[
fontsize=\footnotesize,
]{python}
from toise import figures_of_merit
fom = figures_of_merit.GZK({'Fictive-Optical': 15})
fom.benchmark(figures_of_merit.TOT.ul)
\end{minted}

The framework also makes it straightforward to construct a differential limit, where the neutrino flux is only considered in smaller energy ranges. For example, we can construct a limit over single or half decade bins in energy. Note that the limit is returned in units of the Ahlers \textit{et al.} flux \cite{Ahlers:2012rz}.

\begin{figure}
    \centering
        \includegraphics[width=0.8\textwidth]{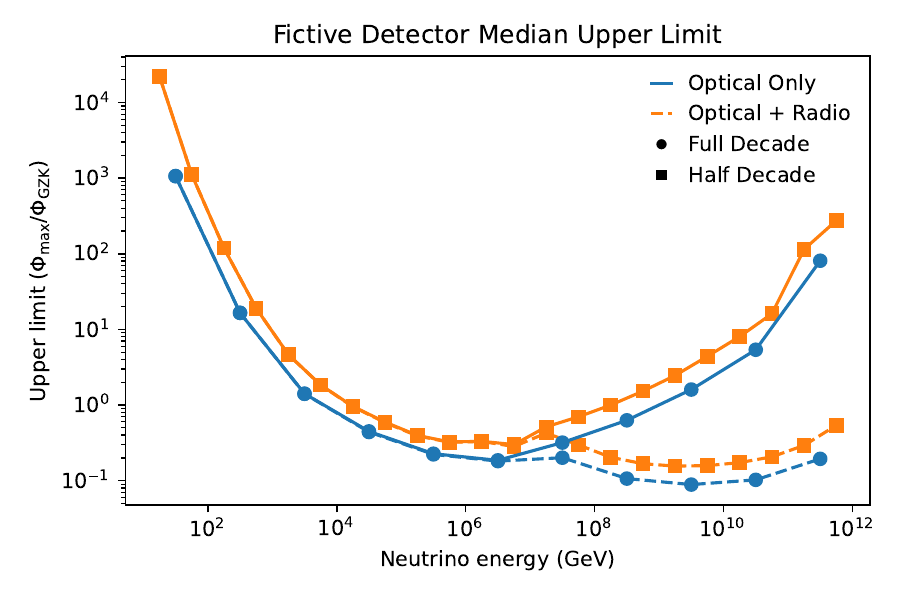}
    \caption{Ultra-High Energy differential flux sensitivity for the fictive detector in units relative to the Ahlers \textit{et al.} cosmogenic neutrino flux \cite{Ahlers:2012rz}, calculated for both half-decade and decade bin widths. The solid lines are for the optical detector alone, while the dashed lines include a fictive radio detector as well. We can see that, especially at high energies, the radio detector has a dramatic contribution to the sensitivity. A major advantage of our framework is that it can easily generate such comparisons and releases the data to be visualized by the end user.}
    \label{fig:uhe_example}
\end{figure}

\begin{minted}[
fontsize=\footnotesize,
]{python}

fulldecade = fom.benchmark(figures_of_merit.DIFF.ul, decades=1)
halfdecade = fom.benchmark(figures_of_merit.DIFF.ul, decades=0.5)
\end{minted}
This results in the detector performance as shown in \autoref{fig:uhe_example}, where both a version with and without a radio array is shown. 

\subsection{Transient (Stacking Search) and Time Integrated Point Source Sensitivity}

In this tutorial, we calculate a sensitivity to a population of transient point sources. In contrast to the diffuse tutorial, this uses full angular and time information, reducing background levels significantly. To start, we build effective areas for the fictive detector configuration.

\begin{minted}[
fontsize=\footnotesize,
]{python}
from toise import factory, diffuse, surface_veto, pointsource, grb, multillh, plotting
aeffs = factory.get('Fictive-Optical')
\end{minted}

\begin{figure}
    \centering
    \includegraphics[width=0.8\textwidth]{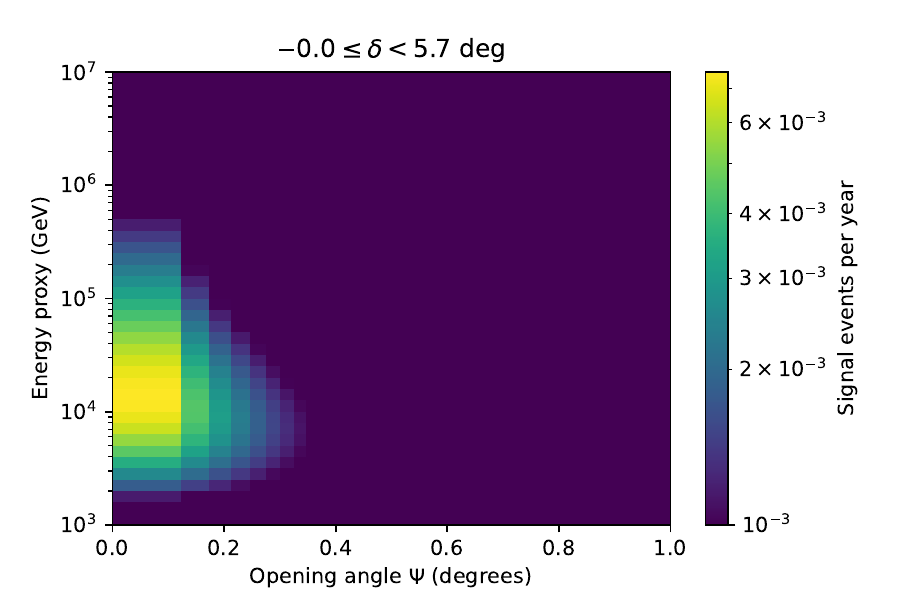}
    \caption{Energy proxy as function of opening angle for a steady point source with neutrino flux of $E^2 \Phi = 10^{-12}$ TeV cm$^{-2}$ s$^{-1}$ located at a declination of \SI{5}{^\circ} as modeled with \fwn{}.}
    \label{fig:expected_counts}
\end{figure}

A targeted point source search is modeled as a counting experiment in bins of opening angle around a known source location, and an energy proxy. To see what this looks like, we can construct the expected counts in these bins from a steady source with a flux of $E^2 \Phi = 10^{-12}$ TeV cm$^{-2}$ s$^{-1}$ located at a declination of 5 degrees. The results are plotted in \autoref{fig:expected_counts}.

\begin{minted}[
fontsize=\footnotesize,
]{python}
from matplotlib import pyplot as plt
from matplotlib.colors import LogNorm
track_aeff = aeffs['unshadowed_tracks'][0]

# find the zenith band corresponding to declination 5 degrees
zi = track_aeff.get_bin_edges('true_zenith_band').searchsorted(-np.sin(np.radians(5)))-1

plt.pcolor(
    np.degrees(track_aeff.get_bin_edges('reco_psi')[:-1]),
    track_aeff.get_bin_edges('reco_energy'),
    pointsource.SteadyPointSource(track_aeff, livetime=1, 
        zenith_bin=zi).expectations(ps_gamma=-2),
    norm=LogNorm(vmin=1e-3))

\end{minted}

\begin{figure}
    \centering
    \includegraphics[width=0.9\textwidth]{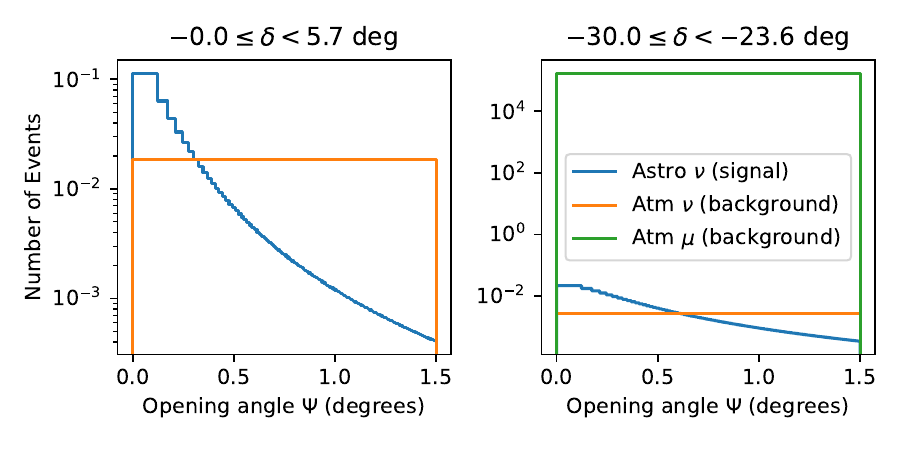}
    \caption{Left: Expected counts from signal GRBs, atmospheric neutrino background, and cosmic-ray-induced muon background just above the horizon for the fictive detector configuration. This assumes a South Pole detector where atmospheric muons are shielded by the Earth's overburden. Right: the same but for a band looking at the sky, therefore showing much higher muon background.}
    \label{fig:S_v_B}
\end{figure}

For a point source search, there is more than signal to worry about. In the Northern sky the background is dominated by atmospheric neutrinos, while in the Southern sky penetrating muons must be considered. The following snippet leads to \autoref{fig:S_v_B}, which illustrates the expected background and signal rates. 

\begin{minted}[
fontsize=\footnotesize,
]{python}
muon_aeff = aeffs['unshadowed_tracks'][1]

for dec in 5, -25:
    zi = track_aeff.get_bin_edges('true_zenith_band').searchsorted(-np.sin(np.radians(dec)))-1
    decstr=dec
    plt.figure()
    ax = plt.gca()
    dec = np.degrees(np.arcsin(-track_aeff.get_bin_edges('true_zenith_band')))
    exes = [
        ('signal', pointsource.SteadyPointSource(track_aeff, livetime=1,
        zenith_bin=zi).expectations(ps_gamma=-2).sum(axis=0)),
        ('neutrino background', diffuse.AtmosphericNu.conventional(track_aeff,
        livetime=1).point_source_background(zenith_index=zi).expectations.sum(axis=0)),
        ('muon background', surface_veto.MuonBundleBackground(muon_aeff,
        livetime=1).point_source_background(zenith_index=zi,
        psi_bins=track_aeff.get_bin_edges('reco_psi')[:-1]).expectations.sum(axis=0)),
    ]
    for label, ex in exes:
        plt.plot(*plotting.stepped_path(np.degrees(track_aeff.get_bin_edges('reco_psi')[:-1]),
            ex),label=label)

\end{minted}

A stacking search is a weighted sum of these sources and backgrounds over all zenith bands. The background expectations are scaled by the same factor, and all event-count expectations multiplied by the assumed livetime. A stacked transient search is similar in \fwn{}, though it works directly with fluences instead of integrating fluxes over an assumed livetime. To model the varying observation time, we can either assume the observation lasts a certain duration (in a mock study of GRBs, this might might be $t_{90}$~\cite{1993ApJ...413L.101K}, the duration over which a GRBs counts rise from 5\% to 95\% of the total) and integrate the backgrounds for that duration, or assume a distribution of integration times, and scale the signal event counts to account for the fraction of the total fluence expected during each observation window. For simplicity in this example we choose the first option.

First, we define a factory function to create predictions for each of our effective areas:

\begin{minted}[
fontsize=\footnotesize,
]{python}
def make_components(aeffs, z=2, t90=45.1, Eiso=53.5, nsources=300):
    aeff, muon_aeff = aeffs
    energy_threshold = np.inf
    atmo = diffuse.AtmosphericNu.conventional(aeff, 1., hard_veto_threshold=energy_threshold)
    atmo.uncertainty = 0.1
    prompt = diffuse.AtmosphericNu.prompt(aeff, 1., hard_veto_threshold=energy_threshold)
    prompt.min = 0.5
    prompt.max = 3
    astro = diffuse.DiffuseAstro(aeff, 1.)
    astro.seed = 2
    zi = slice(None) # use all zenith bands
    livetime = t90*nsources # single burst duration
    atmo_bkg = atmo.point_source_background(zenith_index=zi, livetime=livetime)
    prompt_bkg = prompt.point_source_background(zenith_index=zi, livetime=livetime)
    astro_bkg = astro.point_source_background(zenith_index=zi, livetime=livetime)

    z = z*np.ones(nsources)
    # assume all sources have the same luminosity and are at the same redshift
    ps = grb.GRBPopulation(aeff,
                           z*np.ones(nsources),
                           Eiso=10**(Eiso*np.ones(nsources)),
                          )
    
    components = dict(atmo=atmo_bkg, prompt=prompt_bkg, astro=astro_bkg, ps=ps)
    if muon_aeff is not None:
        components['muon'] = surface_veto.MuonBundleBackground(muon_aeff,
        1).point_source_background(zenith_index=zi,psi_bins=aeff.get_bin_edges('reco_psi')[:-1],
        livetime=livetime)
    return components

# Assume a 15-year exposure. 
# Essentially, assume you see copies of the same 300 bursts every year.
bundle = factory.component_bundle({'Fictive-Optical': 15},
    partial(make_components, z=2, t90=45.1, Eiso=53.5))
\end{minted}
Using this, we fit for the median model discovery factor (returned from this segment as \texttt{mdf}):
\begin{minted}[
fontsize=\footnotesize,
]{python}
components = bundle.get_components()
ps = components.pop('ps')
components['gamma'] =  multillh.NuisanceParam(-2.3, 0.5, min=-2.7, max=-1.7)
components['ps_gamma'] =  multillh.NuisanceParam(-2, 0.5, min=-2.7, max=-1.7)
kwargs = {k:v.seed for k,v in components.items()}
mdf, ns, nb = pointsource.discovery_potential(ps, components, **kwargs)
> INFO:root:baseline: 1.2e+03 actual 8.4 ns: 3.4 nb: 9.3e+05 ts: 25
\end{minted}

\begin{figure}
    \centering
    \includegraphics[width=0.8\textwidth]{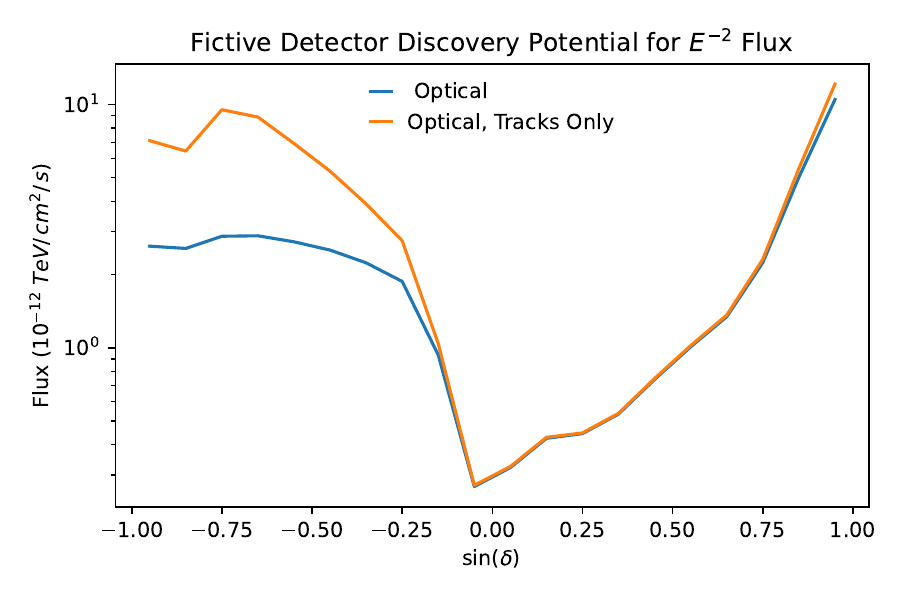}
    \caption{Here we show the 5$\sigma$ discovery potential of the fictive optical detector, in one version including cascades and tracks and in the other including only throughgoing tracks.}
    \label{fig:ps_disco}
\end{figure}

So, for the fictive detector configuration, 50\% of 15-year exposures of the fictive detector would expect to see a 5 sigma excess (TS = 25) over isotropic background by collecting events for 45 seconds around the time of the presumed 300 bursts per year. To adapt this to another model, we would need to change the way the total fluence is calculated for each zenith band. This includes:
\begin{itemize}
\item the overall normalization
\item the energy dependence of the differential fluence from each transient
\item the number of transients per year
\item the $t_{90}$ of each transient
\end{itemize}
As discussed in \autoref{subsec:uhe_ex}, there is also a \texttt{figures\_of\_merit} approach for some of the point-source-related detector benchmarks. In the loop below, we use this approach to easily calculate the discovery potential for the fictive detector. For this example, we compare a throughgoing track search with an all-events (tracks and cascades) search. 
\begin{minted}[
fontsize=\footnotesize,
]{python}

from toise import figures_of_merit 
bins=len(factory.default_cos_theta_bins)
outdat1=np.zeros((bins,3))
outdat2=np.zeros((bins,3))

for i,zi in enumerate(factory.default_cos_theta_bins[:-1]):
    #set up separate discovery potential calculations for Fictive-Optical
    discpot=figures_of_merit.PointSource({'Fictive-Optical': 15}, i) 
    #run a separate FOM class for each detector, 
    #adding multiple to the same bundle combines likelihoods
    outdat1[i]=discpot.benchmark(figures_of_merit.TOT.ul)
    #set up separate discovery potential calculations for Fictive-Optical-TracksOnly

    discpot=figures_of_merit.PointSource({'Fictive-Optical-TracksOnly':15}, i)
    outdat2[i]=discpot.benchmark(figures_of_merit.TOT.ul)

\end{minted}
One can then proceed to plotting the results, as shown in \autoref{fig:ps_disco} using:
\begin{minted}[
fontsize=\footnotesize,
]{python}

midpoints=(factory.default_cos_theta_bins[:-1]+factory.default_cos_theta_bins[1:])/2
plt.plot(midpoints,outdat1[:-1,0],label="Fictive Optical")
\end{minted}

\subsection{Effective Areas}
The third notebook covers the details on how to directly plot the effective area for varying detector configurations using the fictive radio and optical detectors as an example. We initialize the \fwn{} factory and use the provided fictive detectors. 

\begin{minted}[
fontsize=\footnotesize,
]{python}
from toise import factory
factory.set_kwargs(psi_bins={k: [0, np.pi] for k in ('tracks', 'cascades', 'radio')})
radio_aeff = factory.get('Fictive-Radio')['radio_events'][0]
optical_aeff = factory.get('Fictive-Optical')['cascades'][0]
\end{minted}

\begin{figure}
    \centering
    \includegraphics[width=0.8\textwidth]{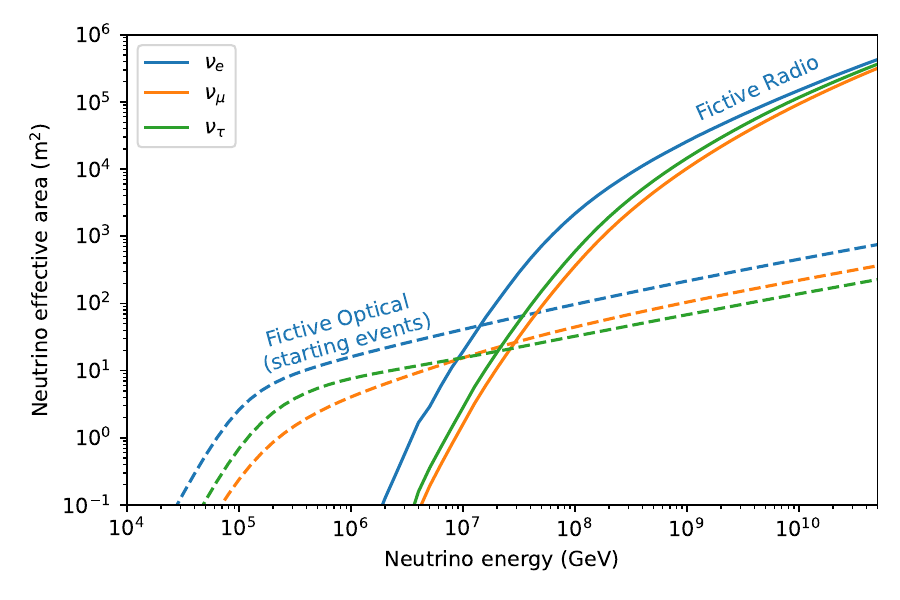}
    \caption{Plotted here are the all-sky averaged effective areas as a function of energy for the fictive detectors. The framework calculates this effective area as a first step to delivering figures-of-merit for the detectors.}
    \label{fig:effAreaEx}
\end{figure}

And then plot the effective areas provided as sum over the whole zenith band, which results (in the full example) in \autoref{fig:effAreaEx}.

\begin{minted}[
fontsize=\footnotesize,
]{python}
cos_theta = radio_aeff.bin_edges[radio_aeff.dimensions.index('true_zenith_band')-1]
flavors = ['${}$' .format([r'\nu', r'\overline{\nu}'][i % 2] + 
            '_{' + ['e', r'\mu', r'\tau'][i//2] +'}') for i in range(6)]

fig = plt.figure()
ax = fig.subplots()
for i in range(0,6,2):
    line = ax.loglog(radio_aeff.bin_edges[0][1:],
        radio_aeff.values[i,...].mean(axis=1).sum(axis=(1,2)))[0]
    line.set_label(flavors[i])
    line = ax.loglog(optical_aeff.bin_edges[0][1:],
        optical_aeff.values[i,...].mean(axis=1).sum(axis=(1,2)), 
        color=line.get_color(), ls='--')[0]

\end{minted}

\subsection{Radio Example}
\label{sec:radio_ex}
The fourth example presents a radio specific analysis, where we evaluate the distribution of detected events for different assumptions about the fictive instrument's performance. For the analysis, we choose a livetime of \texttt{nyears} years, and a fictive array of \texttt{nstations} stations. 
We also define the direction resolution binning. This is the last dimension of the toise effective area tuples. Setting a lower \texttt{psi\_max\_rad} will exclude a fraction of the angular resolution CDF above \texttt{psi\_max\_deg} from analysis. 
For a diffuse analysis we will not constrain it for now, and also choose only a single bin.

\begin{minted}[
fontsize=\footnotesize,
]{python}
nyears = 10
nstations = 30
psi_max_rad = np.pi
psi_bins = np.sqrt(np.linspace(0, psi_max_rad**2, 1))
\end{minted}

We then generate a radio effective area with the resolution parameters for our fictive detector defined in a .yaml file. The parameters can also be updated by hand.
\begin{minted}[
fontsize=\footnotesize,
]{python}
from toise import radio_aeff_generation
radio_array = radio_aeff_generation.radio_aeff('radio_config.yaml', psi_bins=psi_bins)
radio_array.set_parameter('detector_setup', 'nstations', nstations)
radio_array.switch_analysis_efficiency(True)
radio_array.switch_energy_resolution(True)
\end{minted}

We now create the \fwn{} effective area tuples. A atmospheric muon background may also be created. As discussed in \autoref{sec:backgrounds}, the normalization of this background at high energies is still activeld debated in the community, so only a simple version of it is provided in \fwn.

\begin{minted}[
fontsize=\footnotesize,
]{python}
radio_aeff = radio_array.create()
backround_muons_aeff = radio_array.create_muon_background()
# Adding to toise framework
factory.add_aeffs("radio_aeff", (radio_aeff, backround_muons_aeff))
\end{minted}

Next, we generate additional effective areas at trigger-level (i.e.\ we do not account tighter analysis selection) and also assume perfect energy resolution, to compare with the more realistic default \texttt{radio\_aeff}. This also demonstrates various functions in the radio classes, such as the turning on and off of analysis efficiencies and the addition of effective areas to the factory.

\begin{minted}[
fontsize=\footnotesize,
]{python}
radio_array.switch_analysis_efficiency(False)
radio_array.switch_energy_resolution(False)
radio_aeff_perfect_E_eff = radio_array.create()
factory.add_aeffs("radio_aeff_perfect_E_eff", (radio_aeff_perfect_E_eff, None))
\end{minted}

\begin{figure}
    \centering
    \includegraphics[width=0.8\textwidth]{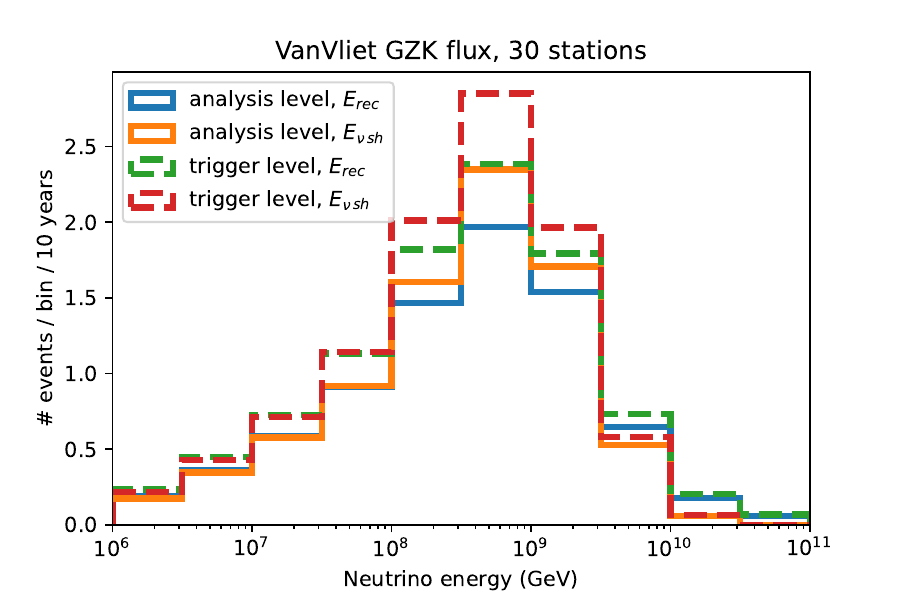}
    \caption{Number of observed events by the fictive radio array of 30 stations, at both the trigger and analysis level, as function of energy for a flux of cosmogenic neutrinos \cite{vanVliet:2019nse}.}
    \label{fig:radio_example}
\end{figure}

The effective areas we have generated above can now be used to evaluate and compare the distributions of events for different flux assumptions. Within the framework, diffuse neutrino fluxes are provided in \texttt{toise.diffuse}. For convenience, the flux plots are provided as functions in this tutorial online and may be fed with any set of previously defined effective area tensors. The essence is provided here: 

\begin{minted}[
fontsize=\footnotesize,
]{python}
from toise import diffuse

def gzk_flux_plot(gzk='VanVliet',aeff_names=["radio_aeff",
                              "radio_aeff_perfect_E_eff"],nyears=1):
    edges = factory.get(aeff_names[0])['radio_events'][0].get_bin_centers('true_energy')
    if gzk == 'VanVliet':
        obj = diffuse.VanVlietGZK
    elif gzk == 'Ahlers':
        obj = diffuse.AhlersGZK
    elif gzk == 'Reasonable':
        obj = diffuse.ReasonableGZK
    else:
        print("ERROR invalid flux")

    for aeff_name in aeff_names:
        astro_w = obj(
            factory.get(aeff_name)['radio_events'][0],
            nyears,
        ).expectations().sum(0)
        (n, bins, patches) = plt.hist(edges, bins=edges, weights=astro_w, histtype='step')

\end{minted}
This may result in the distributions shown in \autoref{fig:radio_example}, illustrating the numbers of events obtained with the different quality cuts from one particular flux.

\end{document}